\documentclass[10pt,aps,prb,floats,eqsecnum,graphicx,epsfig,euscript,amsmath,amssymb,onecolumn]{revtex4-1}
\usepackage[dvips]{graphicx}
\usepackage{epsfig}
\usepackage[centerlast,small]{caption2}
\DeclareMathAlphabet\EuScript{U}{eus}{m}{n} \SetMathAlphabet\EuScript{bold}{U}{eus}{b}{n}

\renewcommand{\min}{\mathop{\rm min}\nolimits}

\def\lapprox{\,\raise0.4ex\hbox{$<$}\kern-0.8em\lower0.7ex\hbox{$\sim$}\,}
\def\gapprox{\,\raise0.4ex\hbox{$>$}\kern-0.8em\lower0.7ex\hbox{$\sim$}\,}
\begin{document}
\bibliographystyle{prsty}
\title{Magneto-phonon resonance in photoluminescence excitation spectra of magneto-excitons in  GaAs/AlGaAs Superlattices}

\author{$\qquad$ S. Dickmann$^{1}$, A.I. Tartakovskii$^{1}$, V.B. Timofeev$^{1}$, $\qquad$ V.M.~Zhilin$^{1}$, J. Zeman$^{2}$, G. Martinez$^{2}$,}

\author{ J.M. Hvam$^{3}$}

\affiliation{$^{1}$Institute of Solid State Physics, Russian
Academy of Sciences, Chernogolovka, 142432 Russia.\\
$^{2}$Grenoble High Magnetic Field Laboratory,
MPI-FKF and CNRS, 38042 Grenoble, Cedex, France.\\
$^{3}$Microelectronic Centre, The Technical University of
Denmark, DK 2800 Lyngby, Denmark}

\begin{abstract}
\vspace{0.mm}
Strong increase in the intensity of the peaks of
excited magneto-exciton (ME) states in the photoluminescence excitation
(PLE) spectra recorded for the ground heavy-hole magneto-excitons
(of the 1sHH type) has been found in a GaAs/AlGaAs superlattice in
strong magnetic field B applied normal to the sample layers. While
varying B the intensities of the PLE peaks have been measured as functions
of energy separation $\Delta E$ between excited ME peaks and the ground state
of the system. The resonance profiles have been found to have maxima at
$\Delta E_{\mbox{\scriptsize max}}$ close to the energy of the GaAs LO-phonon.
However, the value of $\Delta E_{\mbox{\scriptsize max}}$ depends on quantum
numbers of the excited ME state. The revealed very low quantum efficiency of
the investigated sample allows us to ascribe the observed resonance to the
enhancement of the non-radiative magneto-exciton relaxation rate arising due to
LO-phonon emission. The presented theoretical model, being in a good agreement
with experimental observations, provides a method to extract 1sHH
magneto-exciton ``in-plane" dispersion from the dependence of
$\Delta E_{\mbox{\scriptsize max}}$ on the excited ME state quantum numbers.

\end{abstract}
\maketitle

\bibliographystyle{prsty}

\section{Introduction}

Optical transitions in semiconductor superlattices (SLs) have received
great attention recently. In contrast to the quasi
two-dimensional (2D) case of quantum wells (QWs), quasi-particles in SLs
are not fully confined in the growth direction $\hat{z}$. Moreover,
periodicity in the distribution of the materials with
different band gaps and elastic constants leads to
the formation of minibands in the case of free carriers
and excitons \cite{ba88,iv95}, and the appearance of additional
optical and acoustic phonon modes \cite{iv95,ju84,so85,ts89,me89,sap95,pu95}.
Numerous investigations concerning optical observation of
excitons and magnetoexcitons (MEs) in QWs (e.g. see Refs. \onlinecite{ba82,gr84,vi90,ba98}) and in Sls \cite{fu94,mi97,ta97}
usually leave aside the questions
related to the non-radiative excitonic relaxation. At the same time,
non-radiative excitonic transitions such as phonon emission
or absorption allow one to probe indirectly the exciton-energy dispersion
${\cal E}(\mbox{\boldmath $k$})$ (\mbox{\boldmath $k$} is the exciton
wave vector). Furthermore, studies of these processes may provide
the only way of the experimental measurement of the excitonic dispersion.
Indeed, such a powerful method as the hot-luminescence technique (reported
for the first time in Ref. \onlinecite{mirl81}), which was widely used
for measurements of the hole band-structure in bulk
GaAs~{} \cite{ruf,kash93} and in  GaAs/AlGaAs QWs \cite{ruf,kashzach}, is
inefficient for the study of the excitonic dispersion. Although excited ME
states were observed in hot magneto-luminescence measurements \cite{mes90},
such experiments can hardly reveal any information on the function
${\cal E}(\mbox{\boldmath $k$})$. This is because hot-luminescence measurements
can only probe excitons in their radiative states, i.e. when $k$ is very
small, so that ${\cal E}(\mbox{\boldmath $k$})\approx {\cal E}(0)$.

The theoretical investigations of excitonic dispersion have attracted large
efforts ever since Gor'kov and
Dzyaloshinskii's work devoted to three-dimensional ME~{}\cite{go68}.
Later, 2D ME was studied in the paper by Lerner and Lozovik
\cite{le80} and also in the works concerning 2D excitons without magnetic
field \cite{ba82,gr84,br85}. Exciton dispersion relations in SLs
presenting the dependencies ${\cal E}(\mbox{\boldmath $k$})$ where
$\mbox{\boldmath $k$}\parallel \hat{z}$
(i. e. minibands) were calculated in Ref. \onlinecite{di90}.
In parallel, the excitonic binding energy in SLs were studied
theoretically and experimentally in Ref. \onlinecite{ch87} (see also Refs.
\onlinecite{chu89,ch89}), and later the binding energies of ground and excited
states in SLs were calculated in magnetic and electric fields \cite{wh90,di91}.
It is worth noting that two-dimensionality and
strong magnetic fields are the features which usually allow the
separation of transverse variables $x$ and $y$ from the longitudinal
one $z$ ($\mbox{\boldmath $B$}\parallel\hat{z}$) and in addition permit
a simplification of the model for valence band due to the removal of degeneracy
\cite{ba88,iv95}.

The present paper is the result of experimental and theoretical
studies on the observation of a resonant behaviour in relaxation
of MEs in type-I GaAs/Al${}_{0.3}$Ga${}_{0.7}$As SLs in a high magnetic
field  perpendicular to the SL layers. In our experiments we detected only
photoluminescence (PL) signal from the ground heavy hole exciton state 1sHH
(we employ the usual notation for 2D exciton states \cite{ba98,ta97})
while the energy of the laser excitation was continuously varied in a
range of ~80 meV above the 1sHH PL peak. At particular magnitudes
of magnetic field we have observed very strong resonant increase of
intensity of peaks corresponding to the excited ME states in PLE spectra.
We interpret this effect as a manifestation of the magneto-phonon resonance
when strong relaxation of the excited ME state takes place via
longitudinal-optic-(LO-) phonon emission. This occurs when frequency
$\omega_{\mbox{\scriptsize LO}}$ is equal to an approximate multiple of
the excitonic cyclotron frequency $\omega_c$ (a similar effect for free
electrons in heterojunctions is reported  in Refs.\onlinecite{ha89,ba91,va96}):
$$
  \hbar\omega_{\mbox{\scriptsize LO}}=N\hbar\omega_c,\quad N=1,2,3,...
                                                       \eqno (1.1)
$$
Here $\omega_c=eB/\mu c$, where $\mu$ is the reduced excitonic mass for
transverse motion in the layer plane. To our best knowledge this is the
first observation of the magneto-phonon resonance for excitons.

The effect has been found only in one SL sample of the investigated
series of SL structures. According to our measurements, the quantum
efficiency of this sample is about two orders of magnitude less
than that for the other investigated structures. This fact
allows us to treat the considered phenomenon as a feature of the enhanced
relaxation of the excited ME states rather than the peculiarity of light
absorption.

The observed resonance profiles (i.e. integrated intensity of enhanced
PLE peaks as a function of their energy separation $\Delta E$ from the
ground state) were found to be rather broad (with half-widths above $5\,$meV)
and have strong maxima at $\Delta E_{\mbox{\scriptsize max}}$ dependent
on the quantum numbers of excited ME states. These facts indicate that
transitions to the ground ME state occur via intermediate non-radiative
states of the ground excitonic 1sHH band.  The theoretical treatment
developed in our report allows us to extract certain information about
the exciton energy dependence on ``in-plane" wave-vector component
$\mbox{\boldmath $q$}$ from the experimental data.

After the description of the experimental results in Section II we
present the theoretical study of the phenomenon in Section III. Finally,
in the discussion of Section IV we demonstrate the comparison of the PLE
data with the theoretical results and present dispersion curves for
the ``transverse" 1sHH band
$\epsilon(q) \equiv{\cal E}(\mbox{\boldmath $q$},k_z=0)$ extracted from
our experiment.

\section{Experiment}

Samples used in our investigations are molecular beam
epitaxy grown type-I GaAs/AlGaAs SLs. The structures
are not intentionally doped and the flat band regime is
realized. All the samples have $L_{w}=8\,$nm wide GaAs QWs, while
the width of AlGaAs barrier is varied from sample to sample
as $L_{b}=2, 3, 5, 10, 20\,$nm. All the SLs consist
of 20 periods $d=L_{b}+L_{w}$.

The PLE experiments were performed in a He cryostat
with a superconducting magnet providing a magnetic field up to
23 T normal to the SL layers. A technique based on employing optical
fibers was used for the sample excitation and collection of the PLE
signal. In the geometry of our experiment the incident laser light
propagates in the direction close to the normal to the SLs layers.
In order to measure the PLE spectra we tuned a double 1 m monochromator
slightly lower or directly to the ground heavy-hole exciton photoluminescence
(PL) peak and scanned the excitation energy of the Ar$^{+}$-pumped
Ti-Sapphire laser. The main features of the PLE spectra remained
unchanged when the detection position was moved in the limits
of the linewidth around the 1sHH PL peak. PLE of two different polarizations,
$\sigma^{+}$ and $\sigma^{-}$, was detected by a cooled GaAs detector
in the photon counting regime.

Fig. 1 displays  typical heavy-hole exciton PLE spectra
for the SL with $L_w=8\,$nm and $L_b=3\,$nm (referred to as sample 8/3
below) recorded in magnetic field applied normal to the SL layers. We present
here the data for $\sigma^+$
polarization only, since the observed behaviour is very similar to that
for $\sigma^-$ polarization. At $B=0$ the most pronounced peaks in the
spectrum are the direct heavy hole 1sHH-exciton (its
position is not clearly resolved in the shown series of spectra
because the laser wavelength was scanned from the high energy
side of the 1sHH line), the indirect heavy hole $I$(1sHH)-exciton
at $1.586\,$eV~ \cite{ta97} and the light hole 1sLH-exciton at
$1.5955\,$eV~ \cite{1sHH}. When $B$ is increased to 6 T, new features
become clearly resolved in the spectra above the 1sLH peak.
Energies of these excited states rapidly increase with $B$. The
origin of the new peaks can be revealed with the help of an elaborate
theoretical analysis of the magneto-exciton band structure.
However, this task lies beyond the scope of our investigation.
In order to understand the nature of the strongest PLE peaks
we carried out a simplified analysis of PLE spectra recorded at
high magnetic fields. As a result it has been found that the peaks
plotted in Fig. 1 by thick lines correspond to MEs formed by electrons
and heavy holes from Landau levels with equal $N=2, 3$~
(2sHH, 3sHH)~{}\cite{ta97}. In what follows we restrict our
investigation to the study of the resonant behaviour of these
ME states.

As it is seen in Fig. 1, already starting from $B$
of several Teslas the energies of 2sHH and 3sHH ME
peaks increase quasi-linearly with $B$. At $B<6\,$T the intensity of
the 2sHH ME peak is weak. Then starting from $B=12\,$T the intensity
of the 2sHH peak grows rapidly, reaching its maximum at $B=14\,$T and
then decreases more slowly. A similar resonant behaviour is clearly
observed for the 3sHH ME at $B \approx 9\,$T.
Similar resonances have been also observed in $\sigma^-$ polarization,
however they occur at slightly larger $B$: at $B=17\,$T for 2sHH and at
$B=9.5\,$T for 3sHH.

To summarize the PLE data of Fig. 1 (and of the similar series for
$\sigma^-$) the integrated intensities of the 2sHH and 3sHH MEs are
plotted in Figs. 2a,b (black and open squares, respectively) for
both polarizations versus the energy separation $\Delta E$ between
their positions and the location of the 1sHH peak. The energy of 1sHH
peak is extracted from another series of PLE measurements.
The variation of $\Delta E$ with increasing magnetic field occurs
due to the stronger diamagnetic shifts of the excited ME peaks with
respect to that of the 1sHH line (see insets of Figs. 2a,b, where it is
seen that $\Delta E$ changes almost linearly with the magnetic field
both for 2sHH and 3sHH MEs). Figs. 2a,b show that a very strong
increase (by a factor of $10\,-\,20$) of ME peak intensities arises
when $\Delta E\approx 35\,$ and $\approx40\,$ meV for the 2sHH and 3sHH
peaks, respectively. These values are very close to the energy of optical
phonons in GaAs/AlGaAs~{} \cite{ts89}. Note however that the resonance for the 2sHH peak appears at smaller $\Delta E$ than that for the 3sHH
peak. At the same time the resonant
enhancement for the 2s$\to$1s transition naturally occurs in stronger
magnetic field than the resonance for the 3s$\to$1s one.

The following features of the resonances should be noted as well:
(i) their decay as a function of $\Delta E$ is slower than
their build-up; (ii) a structure is
observed at $\Delta E\approx 37\,$meV for 2sHH ME and at
$\Delta E\approx 45\,$meV for 3sHH state.

As it is mentioned above, the precise comparative experiments
showed that sample 8/3 had quantum efficiency
$\eta$ about two orders of magnitude lower than those of all other
SL samples investigated. Meanwhile, other features characterising the
quality of the samples such as PL linewidths and Stokes shifts in
PLE spectra are very similar for all samples (1.5-2 meV and 1-1.5
meV, respectively). Moreover, the heavy-hole exciton binding energy
was found to decrease continuously from the sample with $L_b=20\,$nm
to the sample with $L_b=2\,$nm without any peculiarity for the structure
with $L_b=3\,$nm~{}~{}\cite{ta97}. These two facts imply that the band
structure of sample 8/3 in the energy range close to the energy
of the superlattice ground state is as yet unperturbed. We can suppose
that strong non-radiative recombination channels are most likely
due to deep trapping centers, which originate from native lattice
defects \cite{defects}. However, the nature of non-radiative centers which
is caused by growth procedure details of a particular sample plays
no significant role in our study.

Closing this Section we would like to note that the intensity
of PLE peaks would reflect the absorption efficiency only in the
case of $\eta=1$. This condition
together with the condition of carriers radiative lifetime being long
compared with their relaxation time would permit the excited MEs to
relax into the ME ground state without scattering into other states which
relax later non-radiatively.
On the other hand, in our case of $\eta \ll 1$ the intensity of some
peaks can be resonantly enhanced by the mechanism which strongly reduces
the relaxation time of the ME transition from the excited state to the
radiative ground state and hence decreases to some extent the probability of
non-radiative ME escape.

\section{Theory}

\subsection{Qualitative consideration of the transition
and formulation of the problem}
\label{A}

First note that both states, namely initial $|i\rangle$,
which is 2sHH
or 3sHH, and final $|f_0\rangle=$1sHH have very small exciton
wave-vectors, since $|i\rangle$ arises by virtue of direct light
absorption, whereas $f_0$-relaxation directly provides the optical
PLE signal. More accurately  if
$\mbox{\boldmath $q$}=(k_x,k_y)$ and $k_z$ are the transverse and
longitudinal
components, then for these states we find for the used experimental
geometry that
$$
  q\lesssim 10^4\,\mbox{cm}^{-1},\quad k_z\lesssim 10^5\,\mbox{cm}^{-1}.
                                                        \eqno(3.1)
$$
Here the right sides are determined by the
homogeneity breakdown due to impurities or by momenta of absorbed
and emitted photons. Meanwhile the actual extent
of the exciton wave functions is of the order of $10\,$nm, since,
it is determined by three values: the
magnetic length $l_B=(c\hbar/eB)^{1/2}$, the
effective exciton Bohr radius $a_B=\hbar^2\varepsilon_0/\mu e^2$, and
the period $d$. In the scale of inverse lengths this reads
$$
  k_0\sim 2\pi/l_B, 2\pi/a_B, 2\pi/d \sim 10^6\,-\,10^7\,\mbox{cm}^{-1}.
                                                             \eqno (3.2)
$$
The significant difference between the values (3.1) and (3.2) allows
one to conclude that the considered resonant transition $i\to f_0$ is
indirect one. Indeed, due to momentum conservation, in a direct transition
the emitted optical phonon would have a negligibly
small wave-vector (3.1). As a
result the macroscopic LO polarization field applied to ME may be
considered to be homogeneous, and in the limit $k/k_0\to 0$ the
corresponding transition matrix element would be simply proportional
to $|\langle i|\mbox{\boldmath $r$}|f_0\rangle|$
(with $\mbox{\boldmath $r$}=\mbox{\boldmath $r$}_2-\mbox{\boldmath $r$}_1$
being the difference between electron and hole positions), which turns out
to be equal to zero because of the identical symmetry of the initial
and final states with respect to inversion. Thus we consider
the LO transition not right into $|f_0\rangle$ but initially
into $|f\rangle$ which is a 1sHH exciton state with the wave vector
$k\sim k_0$. Finally, the transition $f\to f_0$ is a non-radiative
process, e.g. provided by acoustic phonon emission.

The assumed transitions are schematically
demonstrated in Fig. 3. The transitions resulting in
non-radiative exciton relaxation are also shown in this diagram.
One can see that in the assumed scheme the PLE signal is
proportional to the rate of the allowed LO phonon emission.

Among the others the diagram reflects one essential
simplification used in our calculations presented below: we
ignore the broadening of the exciton peaks, which naturally
occurs due to crystal and SL imperfections (the curves in
Fig. 3 have zero widths). The diagram implies that
$\Delta E=E_i-E_{f_0}$ should be larger than
$\hbar\omega_{\mbox{\scriptsize LO}}$. However, in the experiment
the beginning of the enhanced relaxation for 2s-exciton occurs
even at $\Delta E \approx 33\,$meV (see Fig. 2), which is lower
than the LO phonon energy in a bulk GaAs crystal $\approx 36\,$meV.
Here we should note that the value of $\Delta E$ (extracted from
PLE spectra) corresponds to the separation of the excited and
ground state exciton absorption maxima. Meanwhile the density
of states in the vicinity of the ground state is presented by a
rather wide band, and at low temperatures the emission comes
mostly from the lower states. This fact provides an effective
Stokes shift; so that the resonant enhancement of the ground-state
luminescence due to LO-phonon mediated relaxation may start at
$\Delta E \approx \hbar \omega_{\mbox{\scriptsize LO}} -
\Delta_{\mbox{\scriptsize PL}} - \Delta_{\mbox{\scriptsize St}}$,
where the PL linewidth $\Delta_{\mbox{\scriptsize PL}} \approx 2\,$meV
and the Stokes shift $\Delta_{\mbox{\scriptsize LO}} \approx 1\,$meV.
Both values $\Delta_{\mbox{\scriptsize PL}}$ and
$\Delta_{\mbox{\scriptsize St}}$ are determined by disorder
effects.

In order to describe the data presented in Fig. 2 in the
approximation of zero exciton-level width we have to employ an
effective energy of LO-phonon which is lower than the tabulated
bulk value. As it was described above this disagreement can be
easily eliminated by the consideration of the finite exciton peak
widths. However, for the simplicity of the model this procedure
is omitted in our calculations. The deviation of the LO-phonon
energy from the bulk value can also be related to an inevitable
effective averaging of the SL multi-mode phonon spectrum in the
single-mode approach employed in our model. Later in Sec. IV we
will briefly discuss the role of the multi-mode spectrum in the
context of the phenomenon studied.

Thus our approach should be considered as a theoretical model,
which simplifies the analytical calculation of the relaxation
rate as follows:

i) results are obtained in the strong magnetic field
approximation, which enables to separate the transverse and
longitudinal variables.

ii) we consider  only the heavy-hole band ignoring its
non-parabolicity and the difference in the
effective hole masses in GaAs and Al${}_{0.3}$Ga${}_{0.7}$As layers,
though we take into account the anisotropy of effective hole mass.

iii) we also ignore the difference of the effective electron masses
in GaAs and Al${}_{0.3}$Ga${}_{0.7}$As layers.

iv) we consider only one LO-phonon mode with the effective energy
$\hbar\omega_0=33\,$meV independent of phonon
wave-vector direction (to avoid the possible misunderstandings we replace
everywhere below $\omega_{\mbox{\scriptsize LO}}$ by $\omega_0$). Besides,
as in the case of bulk GaAs~ \cite{gale87} we use only the Fr\"ohlich-type
Hamiltonian for electron-LO-phonon as well as for hole-LO-phonon
interactions (c.f. Refs.
\onlinecite{iv95} and \onlinecite{na89}).

v) we ignore possible momentum conservation breakdown and finite widths
of the excitonic peaks which occur
due to random impurity potential or quantum well and interface roughness.

vi) finally, any spin-orbit terms in the used Hamiltonians
are disregarded, therefore the presented theory does not take into
consideration the effects of ``fine structure" in the PLE signal dependent
on light polarization (such effects can be rather peculiar, see e.g.
Ref. \onlinecite{ba98}).

In spite of these essential assumptions we believe that
such a simplified approach accounts for the most important aspects
of the transition shown in Fig. 3 and
yields reliable information about the ME relaxation rate.

According to Fig. 3 a considerable enhancement takes place when the
intermediate state $|f\rangle$ is a real (not virtual) state of the lowest
excitonic band. The general formula for the total probability of the LO
phonon emission is
$$
  W_{\mbox{\scriptsize LO}}=\sum_{f} W_{if}\,,    \eqno (3.3)
$$
where $W_{if}$ is the probability of the transition into the
state $|f\rangle$. Meanwhile subsequent processes of relaxation
$f\to f_0$ are not the matter of our interest here.

Evidently the energy conservation leads to the equations
$$
  E_i(B)-E_0(B)= \hbar\omega_0+{\cal E}(\mbox{\boldmath $k$})\,,
  \qquad
  E_f-E_0={\cal E}(\mbox{\boldmath $k$})\,.               \eqno (3.4)
$$
Here the intermediate exciton state energy is written as
$E_{f}=E_0+{\cal E}(\mbox{\boldmath $k$})$ where $E_0=E_{f_0}$ is the
ground state energy and ${\cal E}(\mbox{\boldmath $k$})$ is the
excitonic kinetic energy. The left side in the first Eq. (3.4)
depends quasi-linearly on $B$ (c.f. the insets in Figs. 2a,b)
because $E_i-E_0\equiv
N\hbar\omega_c+\delta U(B)$, where $\delta U$, being much less than
$\hbar\omega_c$, is the difference of
binding energies in $|i\rangle$ and $|f_0\rangle$ states.

At the same time we will see that in strong
magnetic field the matrix element for the $i\to f$
transition has a rather sharp maximum in the vicinity of
$q=q_m \sim l_B^{-1}$,
which provides really the resonant
dependence of PLE signal on the magnetic field.

Let us employ the same material
parameters as in Refs. \onlinecite{di91,di90}, i.e. for the
transverse (in the layer of the well) and for the longitudinal
hole masses we get $m^*_{h\perp}=0.18m_e$ and
$m^*_{h\parallel}=0.34m_e$, respectively. The electron mass is
$m^*_{e\perp}=m^*_{e\parallel}=0.067m_e$, and the dielectric
constant $\varepsilon_0$ is equal to 12.5. Then
the ``transverse" excitonic Bohr radius
$a_0=\hbar^2\varepsilon_0/\mu e^2$  is $14\,$nm, and
for the actual magnetic fields we obtain $l_B\le L_w<a_0$. This
fact justifies the employment of the strong magnetic field
approximation:
$
  l_B \ll a=\min{(a_0,\: L_w)}
$
($a$ is the characteristic distance between an electron and a hole in
${\hat z}$-direction).

The specific dependence ${\cal E}(\mbox{\boldmath $k$})$ is
unknown. Nevertheless, the calculations for ``free"
exciton\cite{go68} in a strong magnetic field and for the
exciton in SL with $L_w=L_b$ and $B=0${}~ \cite{di90} make it
possible to estimate this value and to find that
$\mbox{\footnotesize $\partial $}{\cal E}/{\mbox{\footnotesize
  $\partial $}q}
  \sim \mu e^4 l_B^2 q/\varepsilon_0^2\hbar^2 \gg
  \mbox{\footnotesize $\partial $}{\cal E}/
  {\mbox{\footnotesize $\partial $}k_z}$
Indeed according to
Ref. \onlinecite{di90} the miniband width for a SL with
$L_w=L_b=3\,$nm is approximately $6\,$meV. Consequently for our SL
with the same $L_b$ and with $L_w=8\,$nm the miniband width should
be smaller than $1\,$meV~ \cite{foot2}.
In a strong perpendicular magnetic field this value is to be even more
strongly reduced and thus it becomes negligible in comparison
with the expected characteristic energy of dependence $\epsilon(q)$.

Further, the summation in Eq. (3.3) leads to the result which
contains the density of allowed states $|f\rangle$. This value
is inversely proportional to the Jacobian of the change from
variables of integration over phase
space to the integration over $f$-exciton energy and over
$k_z$ wave-vector component:
$$
  \frac{\mbox{\footnotesize $\partial $}({\cal E},\: k_z)}{
  \mbox{\footnotesize $\partial $}(q,\:k_z)}\approx
  d\epsilon/dq\,.                                    \eqno (3.5)
$$

The transition probabilities in Eq. (3.3) are expressed in terms of
the relevant matrix element ${\cal M}_{i\to f}$,
$$
  W_{if}=\frac{2\pi}{\hbar}|{\cal M}_{i\to f}|^2\delta
  (E_i-E_{f}-\hbar\omega_0)\,,                       \eqno (3.6)
$$
which in their turn is calculated using the wave functions of
the excitonic states.

\subsection{Excitonic wave functions}
\label{B}

We can write the excitonic wave functions
in the following manner (c.f. Refs. \onlinecite{go68,le80}):
$$
  \Psi(\mbox{\boldmath $r$}_1,\mbox{\boldmath $r$}_2,z_1,z_2)=
  {\cal L}^{-1}\cdot \exp{\left(i\mbox{\boldmath $R$}\mbox{\boldmath $q$}+
  \frac{i}{2l_B^2}[\mbox{\boldmath $r$}_1\times \mbox{\boldmath $r$}_2]
  \mbox{\boldmath $B$}/B\right)}
  \Phi(\mbox{\boldmath $r$}-\mbox{\boldmath $r$}_0)F(z_1,z_2).
                                                       \eqno (3.7)
$$
Here $j=1,\:2$ denotes electron and hole;
$\mbox{\boldmath $r$}_j=(x_j,y_j)$ is the 2D vector;
$\mbox{\boldmath $r$}=\mbox{\boldmath $r$}_1-\mbox{\boldmath $r$}_2,\quad$
$\mbox{\boldmath $R$}=(\mbox{\boldmath $r$}_1+\mbox{\boldmath
$r$}_2)/2,\quad$ $\mbox{\boldmath $r$}_0=\mbox{\boldmath
$B$}\times\mbox{\boldmath $q$}l_B^2/B;\quad$ ${\cal L}$ is the sample
size in
the plane $({\hat x},\hat{y})$. $\Phi(\mbox{\boldmath $r$})$ obeys
two dimensional Sr\"odinger equation in the main approximation of which
the Coulomb interaction can be neglected \cite{go68,le80}. In this
case
$\Phi(\mbox{\boldmath $r$})\approx |N,m,\mbox{\boldmath $r$}\rangle$
($N$ is the Landau level number,
$m$ is the magnetic quantum number), where
$$
  |N,m,\mbox{\boldmath $r$}\rangle=
  \left[\frac{N!}{2^{|m|+1}(N+|m|)!\pi}\right]^{1/2}
  r^{|m|}l_B^{-|m|-1}L_N^{|m|}(r^2/2l_B^2)e^{im\varphi-r^2/4l_B^2}
                                                         \eqno (3.8)
$$
($L_N^m$ are Laguerre polynomials). The energies corresponding to these
functions are
$$
  E^{(0)}_{Nm}=
  \hbar\omega_c[N+1/2(|m|+\gamma m+1)],
$$
where $\gamma=(m_{h\perp}^*-m_e^*)/(m_{h\perp}^*+m_e^*)$.

In the next approximation Coulomb interaction can be taken into account
with the use of the operator
$$
  {\cal H}_{\mbox{\scriptsize int}}=
  e^2\left/\varepsilon_0\sqrt{
  (\mbox{\boldmath $r$}+\mbox{\boldmath $r$}_0)^2+w^2}\right.,\quad
  \mbox{where}\qquad
   w=z_1-z_2\,,                                            \eqno (3.9)
$$
in perturbation theory\cite{go68,le80}. Coulomb interaction in
function $F(z_1,\,z_2)$ should be included from the very
first step. The Bloch theorem for this function takes place
if one changes the variables; namely if
$$
  {\cal F}_{k_z}(Z,w)= F(Z+\gamma_1 w,\; Z-\gamma_2 w)\,, \quad
  \mbox{where}\quad \gamma_1=m_{h\parallel}^*/M,\quad \gamma_2=m_e^*/M
  \,,                                              \eqno (3.10)
$$
then
$$
  {\cal F}_{k_z}(Z,w)={\cal L}_z^{-1/2}e^{ik_zZ}v_{k_z}(Z,w)\,,\qquad
  \mbox{where}\quad v_{k_z}(Z+d,w)= v_{k_z}(Z,w).  \eqno (3.11)
$$
(${\cal L}_z$ is the sample size along ${\hat z}$). We restrict ourselves
to the one-band approximation assuming for all $i$ and $f$ states that
$v_{k_z}(Z,w)$ presents  the ground state function of the two-particle
motion in $\hat{z}$ direction. This function can evidently be
normalized so that $\int_{-\infty}^{+\infty}\int_{Z_0}^{Z_0+d}
|v_{k_z}(Z,w)|^2dZdw=d$

\subsection{Matrix element calculation and inverse transition
time}
\label{C}

Optical phonons in SLs have considerable energy dependence on
their wave-vector direction. Moreover, for arbitrary direction the
classification of the optical phonon branches as longitudinal and transverse
is impossible due to the inhomogeneity of the superlattice medium along
the $z$-axis.
We choose the simplified model and calculate ${\cal M}_{i\to f}$
employing the Hamiltonian of
exciton-LO-phonon interaction in the following form\cite{na89}:
$$
  {\cal H}_{\mbox{\scriptsize opt}}=\frac{1}{{\cal L}}\left(\frac{\hbar}{{\cal L}_z}\right)^{1/2}
  e^{-i\omega_{\mbox{\tiny 0}}t}\sum_{\mbox{\footnotesize{{\boldmath $k$}}}}
  U_{\mbox{\scriptsize opt}}(\mbox{\boldmath $k$})\left(
  e^{i{\mbox{\footnotesize{{\boldmath $qr$}${}_1$}}}+ik_zz_1} -
  e^{i{\mbox{\footnotesize{{\boldmath $qr$}${}_2$}}}+ik_zz_2}
  \right) + \mbox{H.c.}\,.                     \eqno (3.12)
$$
Final results include only the squared modulus of the vertex, which is
$$
  |U_{\mbox{\scriptsize opt}}|^2=\frac{2\pi e^2\omega_0}{\overline{\varepsilon}k^2}\,.
                                                 \eqno (3.13)
$$
Here ${\overline\varepsilon}^{-1}= \varepsilon^{-1}_{\infty}-
\varepsilon^{-1}_0$ (the standard notations\cite{gale87} are
used).

Note that the calculation of $\langle i|{\cal H}_{\mbox{\scriptsize opt}}|f\rangle$ with the functions (3.8) without Coulomb interaction
gives exactly zero in the result.
Indeed the factorization in the form of the product of one-particle
$N$-Landau-level functions is always possible for these functions.
Therefore if we are interested in the transition between the
different levels $N_1$ and $N_2$, then the matrix element of a
one-particle operator always includes the convolution over the
transverse variables $\mbox{\boldmath $r$}_j$ of one of the
particles $\langle N_1,m|m,N_2\rangle$ which is zero because $N_1\not=N_2$.
Thus ${\cal M}_{i\to f}$ is defined by Coulomb corrections
to $\Phi(\mbox{\boldmath $r$})$, which were discussed above.
Taking into account this comment, the final expression for the
matrix element is
$$
  {\cal M}_{i\to f}=\sum_v\frac{\langle v,\mbox{\boldmath $k$}|
  {}_{\perp}\!\langle f,\mbox{\boldmath $k$}|{\cal H}_{\mbox{\scriptsize int}}|
  v,\mbox{\boldmath $k$}\rangle\!{}_{\perp}|{\cal H}_{\mbox{\scriptsize opt}}|i,
  0\rangle}{E_i^{(0)}-E_v^{(0)}-\hbar\omega_0} +
  \sum_v\frac{\langle f,\mbox{\boldmath $k$}|{\cal H}_{\mbox{\scriptsize opt}}|
  {}_{\perp}\!\langle v,0|{\cal H}_{\mbox{\scriptsize int}}
  |i,0\rangle\!{}_{\perp}|v,0\rangle}
  {E_{f}^{(0)}-E_v^{(0)}+\hbar\omega_0}
  \,.\eqno (3.14)
$$
Here to calculate the expectations one should use the following rules:
Brackets ${}_{\perp}\!\langle...,\mbox{\boldmath $k$}|...|...,
\mbox{\boldmath $k$}\rangle\!{}_{\perp}$
mean the integration over the transverse variables of Coulomb energy
with the functions (3.7); if $\mbox{\boldmath $q$}=0$ then
in Eq. (3.9) $\mbox{\boldmath $r$}{}_0=0$.
Usual brackets $\langle...\rangle$
mean full expectation (the integration over
$\mbox{\boldmath $r$}{}_1, z_1, \mbox{\boldmath $r$}{}_2, z_2$).
Denominators in (3.14) contain the energy values in zero-order
approximation in the interaction and hence one should assume that
$E_i^{(0)}-E_{f}^{(0)} = N\hbar\omega_c \approx
\hbar\omega_0$.
Indexes $i$, $v$ $f$ are two-dimensional (each of them is a pair $N,m$).
In our case $i=(N,0)$, $f=(0,0)$. The allowed $v$ in the first sum are
$(N,\pm m)$ or $[s,\pm (N-s)]$ with $s=0,\,1,...N-1$; and in the
second sum $v=(0,0)$. Note also that for $\mbox{\boldmath $q$}=0$
the value (3.14) becomes zero.

After all the integrals are calculated, the matrix element
to the first order in $l_B/a$ takes the form:
$$
  {\cal M}_{i\to f}=\left(\frac{\hbar}{V}\right)^{1/2}
  \frac{e^2}{\hbar\omega_c\varepsilon_0}U_{\mbox{\scriptsize opt}}
  T(k_z){\cal G}_N(ql_B)\,,\eqno (3.15)
$$
where $V={\cal L}^2{\cal L}_z$,
$$
  T(k_z)=\langle{\cal F}_{k_z}(Z,0)|e^{ik_zZ}|{\cal F}_0(Z,0)\rangle=
  d^{-1}\int_{Z_0}^{Z_0+d}v^*_{k_z}(Z,0)v_0(Z,0)\,dZ\;\sim\;
  \frac{1}{a}\,,                                     \eqno (3.16)
$$
$$
  {\cal G}_1(p)=-\frac{4\gamma}{1-\gamma^2}e^{-p^2/4}(1-e^{-p^2/2})+
  \sum_{m=1}^{\infty}\frac{\gamma e^{-3p^2/4}mp^{2m}(m+1-p^2)}{2^m(m+1)!
  [(1+m/2)^2-m^2\gamma^2/4]}\,,                      \eqno (3.17)
$$
$$
  {\cal G}_{2}(p)=\frac{-\gamma(1-p^2)}{1-\gamma^2}e^{-p^2/4}
  \left[1-e^{-p^2/2}(1+p^2/2)\right]-
  \frac{2\gamma}{9-\gamma^2}p^2(1+p^2/4)e^{-3p^2/4}
$$
$$
  +\sum_{m=1}^{\infty}\frac{\gamma me^{-3p^2/4}(p^2/2)^m
  (m+1-p^2/2)\left[m^2+
  3(1-p^2/2)(m-p^2)+2\right]}{2(m+2)!
  \left[(2+m/2)^2-m^2\gamma^2/4\right]}\,.            \eqno (3.18)
$$

Now we have to substitute the expression (3.15) into Eq. (3.6).
Then in Eq. (3.3) with the help of Eq. (3.5) we change
from the summation over states $f$ to the integration over phase
space and further to the integration over $\epsilon$ and $k_z$. We find
then that the $\delta$-function in Eq. (3.6) removes the
integration over $\epsilon$. Finally,
the result for the total probability of the
transition from the excited ME state to some state of the ground ME band
is
$$
  W_{\mbox{\scriptsize LO}}(B)=\frac{e^6\omega_0\Lambda(q)G_{N}(ql_B)}
  {(\hbar\omega_c)^2\varepsilon_0^2\overline{\varepsilon}
  d\epsilon/dq}\,,                                        \eqno (3.19)
$$
where
$$
  \Lambda(q)=\frac{q}{\pi}\int_{-\infty}^{-\infty}
  \frac{|T|^2dk_z}{q^2+k_z^2}\,,                          \eqno (3.20)
$$
and the functions
$$
  G_N(p)=\pi{\cal G}_N^2(p)                               \eqno (3.21)
$$
are plotted in Fig. 4. One should bear in mind that $q$
in Eq. (3.19) is not independent value but $q=q(B)$
is the root of first Eq. (3.4) with
${\cal E}(\mbox{\boldmath $k$})\approx \epsilon(q)$
Therefore $W_{\mbox{\scriptsize LO}}$ is the function of magnetic field which
may be converted to the function of $(N+1)$sHH-exciton peak position
(see insets in Figs. 2a,b).

Note also that the maximum of function
$G_2$ is shifted substantially to higher $q$ than that of
function $G_1$. This fact accounts for the smaller experimentally
observed $\Delta E_{\mbox{\scriptsize max}}$ for the 2s$\to$1s transition
with respect to that of the 3s$\to$1s one.

Finally one can estimate with the help of Eq. (3.19) the inverse time
of relaxation $i\to f$ in the vicinity of the
resonance when $q\simeq q_{Nm}$.
If $B\simeq 10\,$T,
$d\epsilon/dq\sim 10^{-5}\,$meV$\cdot$cm and $|T|\sim 10^{6}\,$cm${}^{-1}$,
then this time is
$$
  \tau_{N}^{\pm}\sim W_{\mbox{\scriptsize LO}}^{-1}\sim 0.01\,\mbox{ns}\,,
                                                                 \eqno(3.22)
$$
where superscript $+$ or $-$ labels exciton spin quantum
numbers $S_z=\pm 1$ which associated with $\sigma^{\pm}$ luminescence
polarizations. Generally, the considerable spin-orbit coupling manifests
itself in the experimental data, and accordingly we should
label by ``$+$" or ``$-$" all the
quantities $T$, $\Lambda$, $\epsilon$, $q$ (considered in their turn as
the functions
of either $B$ or the peak position $\Delta E=E_i-E_{0})$.

\section{Discussion}
The intensity of
the PLE signal under the resonant conditions should be proportional to
the inverse time (3.22). Nevertheless an
immediate comparison with experimental data of Figs. 2a,b is
impossible as long as the functions $\epsilon_{\pm}(q)$
are unknown. The alternative approach is to find these
functions being guided by this comparison. Note that the following
results for $\epsilon_{\pm}$ obtained below are rather qualitative
and should be considered as an estimate of the energy dependence on the
component $q$.

Let us specify the
form of energy dispersion phenomenologically as
$$
  \epsilon_{\pm}(q)={\cal E}_b\frac{(g^{\pm}_1+
  g^{\pm}_2B^{1/2})(ql_B)^2}
  {1+g^{\pm}_3(ql_B)^2}\,,                          \eqno (4.1)
$$
where ${\cal E}_b$ is a parameter of the order of the exciton binding
energy\cite{ch87,wh90} and $B$ is
measured in Teslas. Naturally the parameters
$g^{\pm}_i$ should be the same
for the both resonant peaks (that is they are independent
of excited exciton quantum number $N$), but the set of these
parameters varies with the spin quantum number.

The functions (4.1) with ${\cal E}_b=5\,$meV are presented
in Fig. 5 for the
specific sets of $g^{\pm}_i$ (indicated in the caption)
and for various
magnetic fields. Now we can find the values $q_{\pm}$ from
Eqs. (3.4) as functions of $\Delta E$ (see Figs. 6a,b). Meanwhile
the dependence $B(\Delta E)$ for magnetic field entering
Eq. (4.1) (and also indirectly through $l_B$) is extracted from
the experiment. Therefore Eqs. (3.4), (3.19) and (4.1) lead to
the formula of relevant PLE intensities $I^{\pm}_N \propto
W_{\mbox{\scriptsize LO}}$, namely, in arbitrary units (a.u.) of Figs. 2a,b:
$$
  I^{\pm}_N(\Delta E)=\frac{C_{\pm}{\cal E}_b\Lambda_{\pm}(q_{\pm})}
  {B^{3/2}(\Delta
  E-\hbar\omega_0)}\left[q_{\pm}l_B+g_3^{\pm}(q_{\pm}l_B)^3\right]
  G_N(q_{\pm}l_B)\,.                       \eqno(4.2)
$$
Here the additional parameters $C_{\pm}$ arise, which are required
for fitting the experimental data presented in arbitrary units.
Besides, one should specify the functions $\Lambda_{\pm}$. We do it
with the help of Eq. (3.20) and two-harmonic expansion for the periodic
functions $T_{\pm}(k_z)$,
$$
  T_{\pm}\propto 1+h_1^{\pm}\cos{(k_zd)}+h_2^{\pm}\cos{(2k_zd)}\,.
                                              \eqno(4.3)
$$
The solid lines in Figs. 2a,b correspond to the dependencies
(4.2) with $C_+=6.1\,$a.u.$\times$T${}^{3/2}$ and
$C_-=10.6\,$a.u.$\times$T${}^{3/2}$. In our calculations we
use $h_1^+=1.68$, $h_2^+=0.902$, $h_1^-=1.79$, $h_2^-=1.34$ and
the sets of functions $g_i^{\pm}$ which are presented in
Fig. 5. These parameters are found during the fitting of the
experimental points of Figs. 2a,b.
Meanwhile the picture of such a comparison remains qualitatively
the same even in the case when $h_1^{\pm}=h_2^{\pm}=0$. It is also
important that all the optimum parameters $C_{\pm},g_i^{\pm}$ and
$h_i^{\pm}$ turn out to be of the
order of $0.1\,-\,1$. This confirms the validity of the
estimate (3.22) and indirectly the choice of the
functions  (4.1) and (4.3).

Summarizing the results of the paper we see
that our theory is in satisfactory agreement with
the experimental data. Comparison with experiment
leads to reasonable dependencies $\epsilon(q)$ which are
presented in Fig. 5. At the same time, a more detailed theory
taking into account real multi-mode phonon
spectrum in SL~{}\cite{iv95,me89} is yet to be developed.
In the real situation we can expect that with changing
$\Delta E$ ME relaxation mediated by different dominant
optic-phonon modes should occur. We think that this change
to another type of optic phonons explains the appearance of the
shoulders in resonance profiles mentioned in Sec. 2. Also one can expect
a quasi-continuous increase of the frequency of emitted phonons in
comparison with the employed parameter $\omega_0=33\,$meV
with increasing $\Delta E$. This implies that the real bands $\epsilon(q)$
are more narrow than those calculated in the frame of our model.
Actually only the initial portions of the Fig. 5 curves [it seems
for $\epsilon(q)<10\,$meV] should reflect a real exciton dispersion.

Finally, note that there are other hypothetical ways for the
excited ME states to increase their intensity in PLE spectra. First,
the increase in the light absorption (not in the ME relaxation) can
be caused by a resonant increase in oscillator strengths of the direct
radiative transition. This can occur due to the mixing of MEs states
with some other quasi-particle states in SL. Second, the increase in the
relaxation rate may appear because the intermediate
$f$-state overlaps ME states of the region of increased density of
states, namely, at SL miniband edge. However, all these opportunities
can not lead to the observed increase in the intensity by a factor of
more than 20 times. Moreover, in these cases similar resonances also
should be observed in the samples with $L_b=2$ and $5\,$nm (of
course at different magnetic fields), which does not occur.

The authors thank V. D. Kulakovskii for useful discussions and
R. M. Stevenson for the critical reading of the manuscript.
This work is supported by Russian Foundation for Basic Research.
A. I. T. and V. B. T. thank also INTAS and the Sci.-Technical
Program on the Physics of Solid State Nanostructures for the support.

\appendix

\begin{figure}[h]
\hspace{-5.mm}
\vspace{-0.mm}\includegraphics[width=1.3\textwidth]{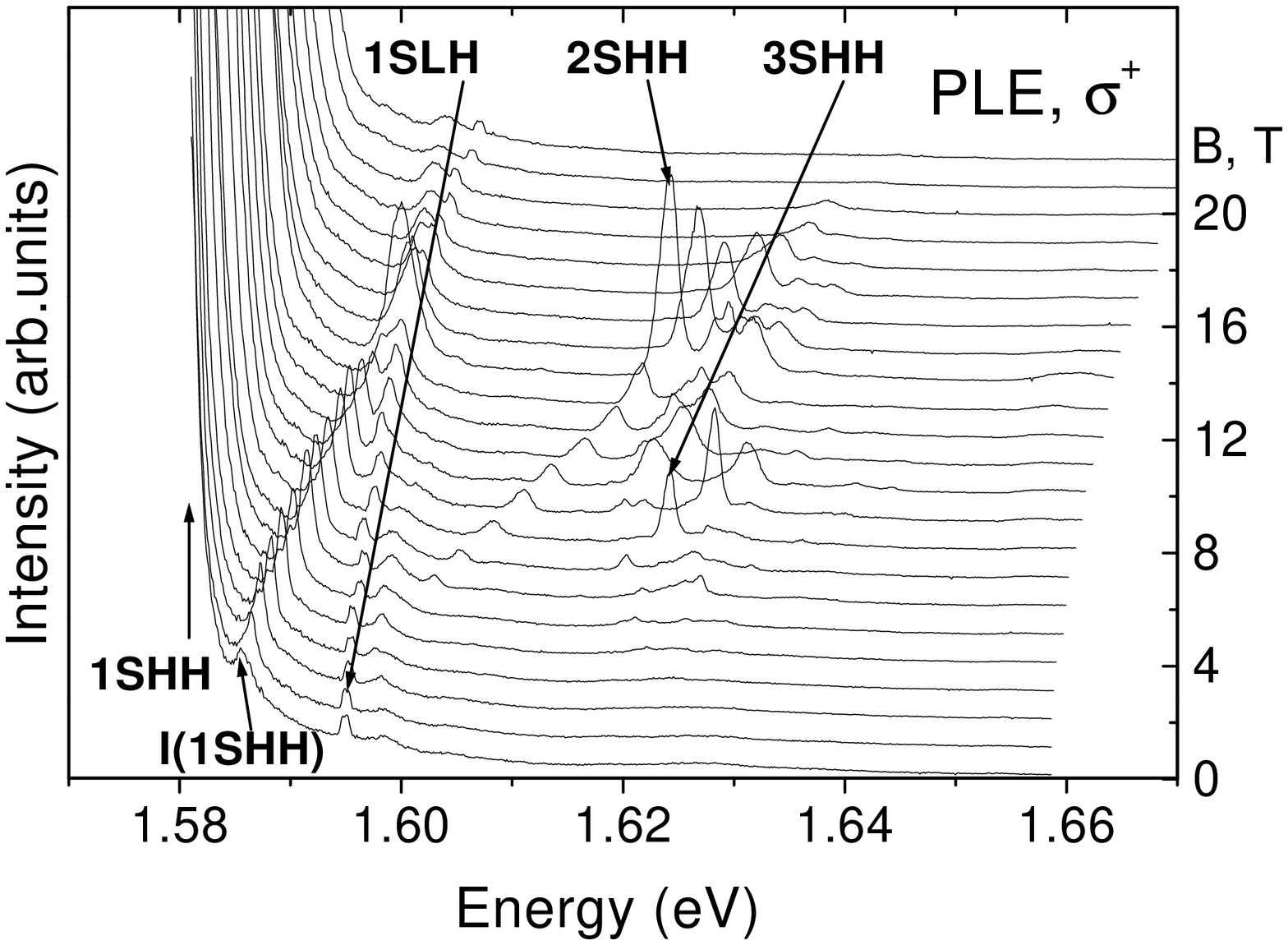}
\vspace{-25.mm}
\caption {\large $\sigma^+$ PLE spectra recorded for ground
magneto-exciton state for various magnetic fields.
The excited magneto-exciton peaks are indicated by arrows.}\vspace{-2mm}
\end{figure}
\begin{figure}
\center \vspace{-0.mm}
\hspace{-5.mm}
\includegraphics[width=.55\textwidth]{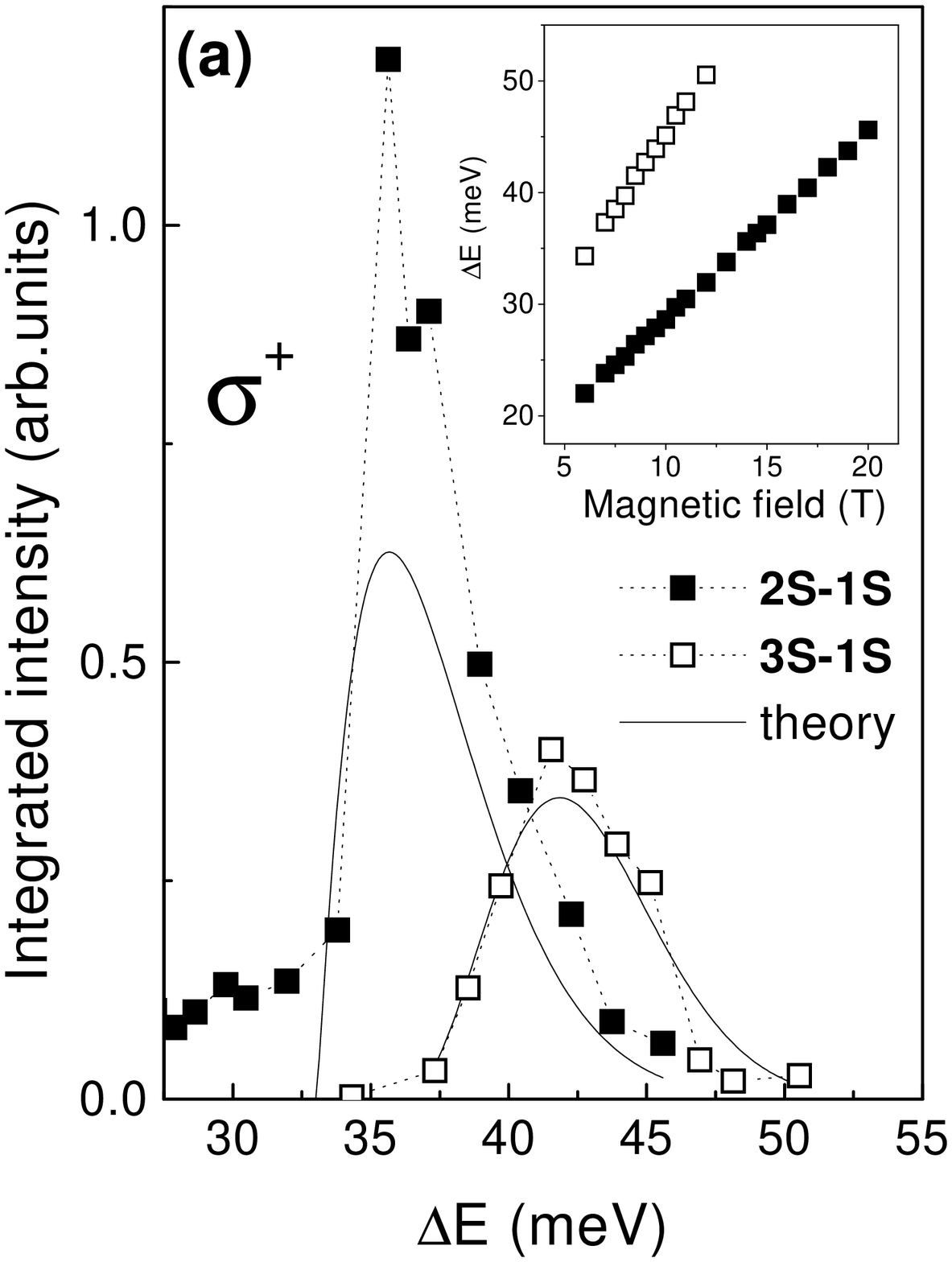} 
\hspace{-14.mm}\includegraphics[width=.55\textwidth]{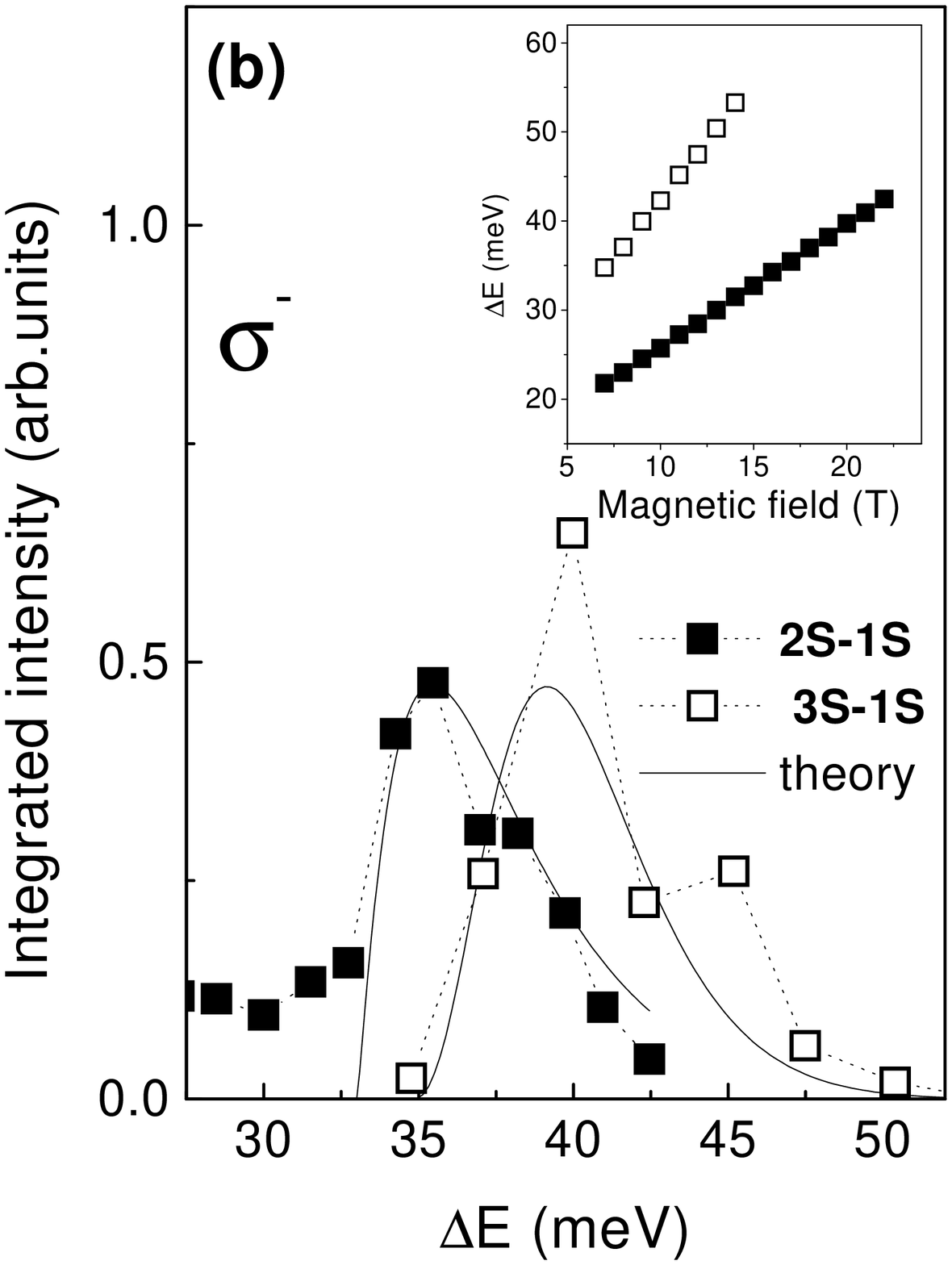}
\vspace{-15.mm}
\caption{\large{Integrated intensities of the 2sHH and 3sHH PLE peaks
for the both polarizations (a~--~$\sigma^+$, b~--~$\sigma^-$)
as a function of the peak
position $\Delta E$ measured from the 1sHH peak energy.
Black and open squares show the experimental data.
The insets present the experimental
dependences of $\Delta E(B)$ for the 2sHH and 3sHH peaks.}}
\end{figure}

\begin{figure}[h]
\begin{center} \vspace{0.mm}
\hspace{-10.mm}
\includegraphics*[width=1.\textwidth]{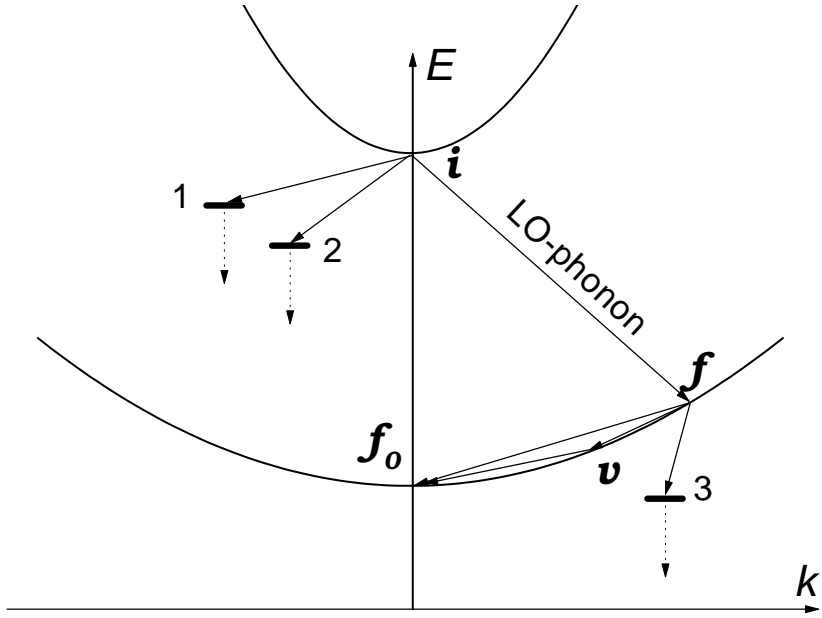}
\end{center}
\vspace{-15.mm}
 \caption{\large Diagram of possible transitions. LO-phonon-emission $i\to f$ gives
rise to the studied effect. Transitions $i\to 1$, $i\to 2$, and
$f\to 3$, lead to nonradiative exciton annihilation.
Transitions $i\to f\to f_0$ and $i\to f\to v\to f_0$
are the examples
of nonradiative 2(or 3)sHH$\to$1sHH relaxation yielding the measured
luminescence signal.}
\end{figure}

\begin{figure}
\begin{center} \vspace{-0.mm}
\hspace{-10.mm}
\includegraphics*[width=1.\textwidth]{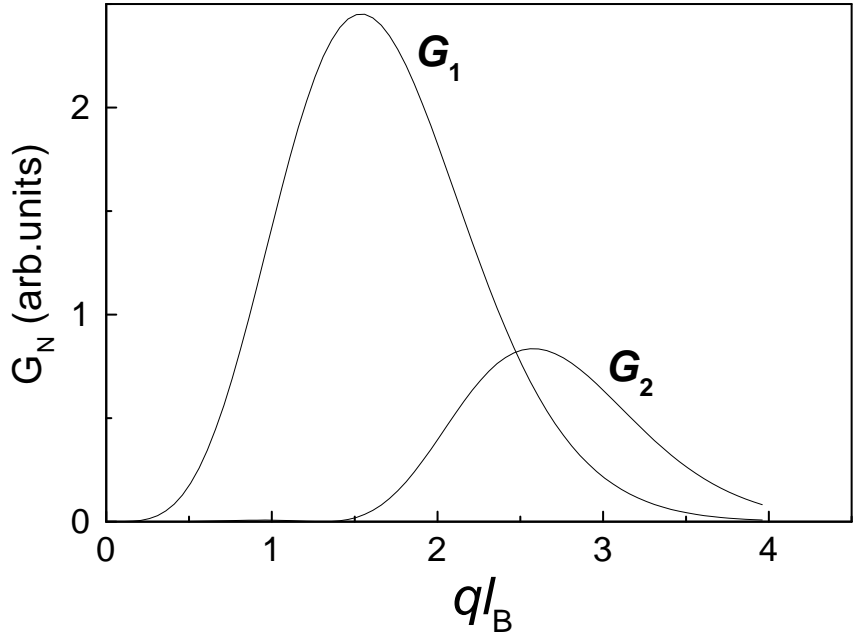}
\end{center}
\vspace{-15.mm}
\caption{\large Functions $G_N$ versus the dimensionless ``in plane"
wave-vector component. Landau level numbers $N=1,\,2$
correspond to 2s$\to$1s and 3s$\to$1s transitions respectively.}
\end{figure}

\begin{figure}[h]
\begin{center} \vspace{-10.mm}
\hspace{-10.mm}
\includegraphics*[width=1.\textwidth]{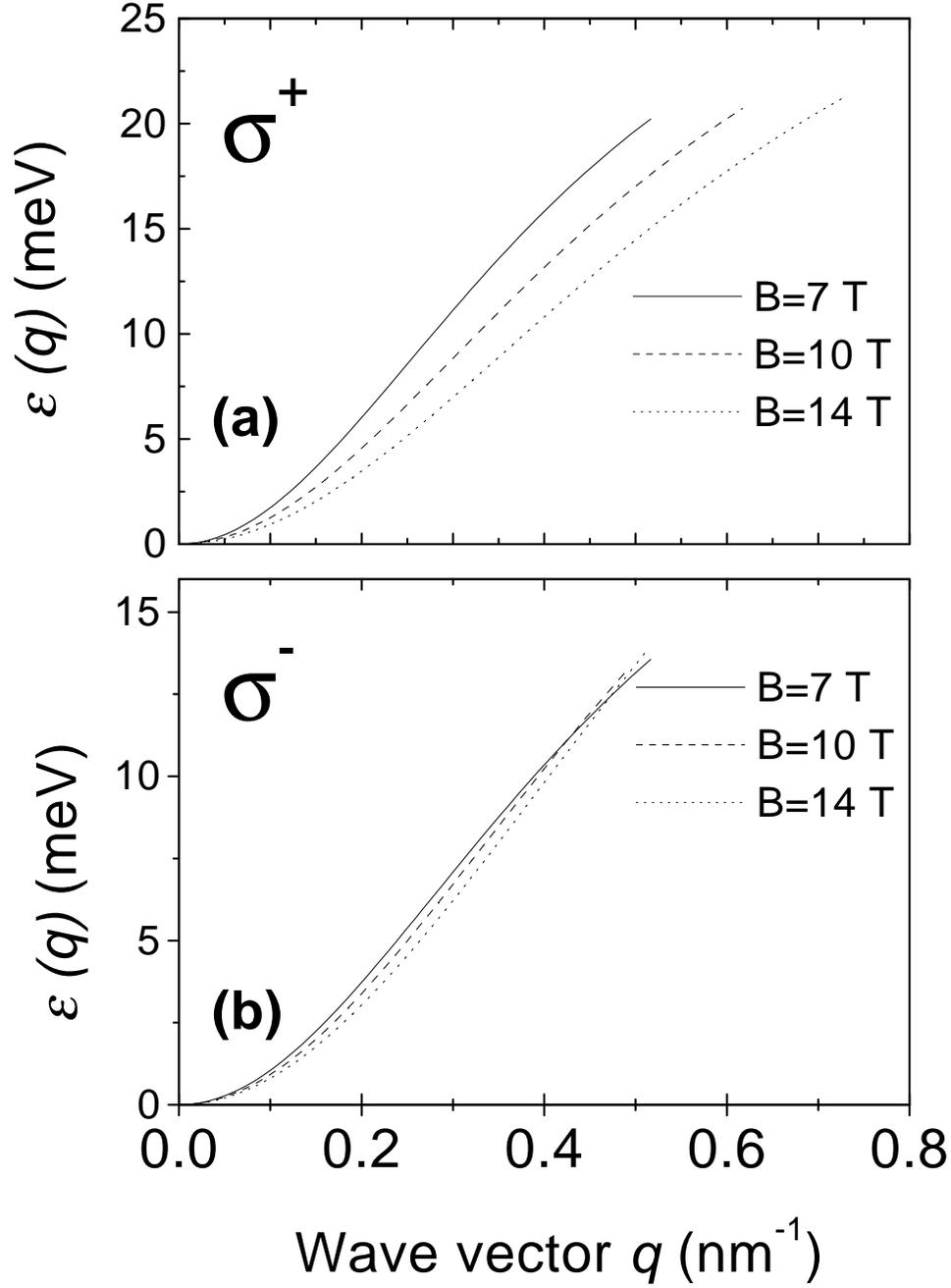}
\end{center}
\vspace{-25.mm}
 \caption{\large Transverse energy dispersion functions of Eq. (4.1) when
${\cal E}_b=5\,$meV;
$g_1^+=0.338$, $g_2^+=0.019$, $g_3^+=0.0561$, for $\sigma^+$-polarization
(a), and
$g_1^-=-0.054$, $g_2^-=0.108$, $g_3^-=0.0459$, for $\sigma^-$-polarization
(b).}
\end{figure}

\begin{figure}[h]
\begin{center} \vspace{-10.mm}
\hspace{-10.mm}
\includegraphics*[width=1.\textwidth]{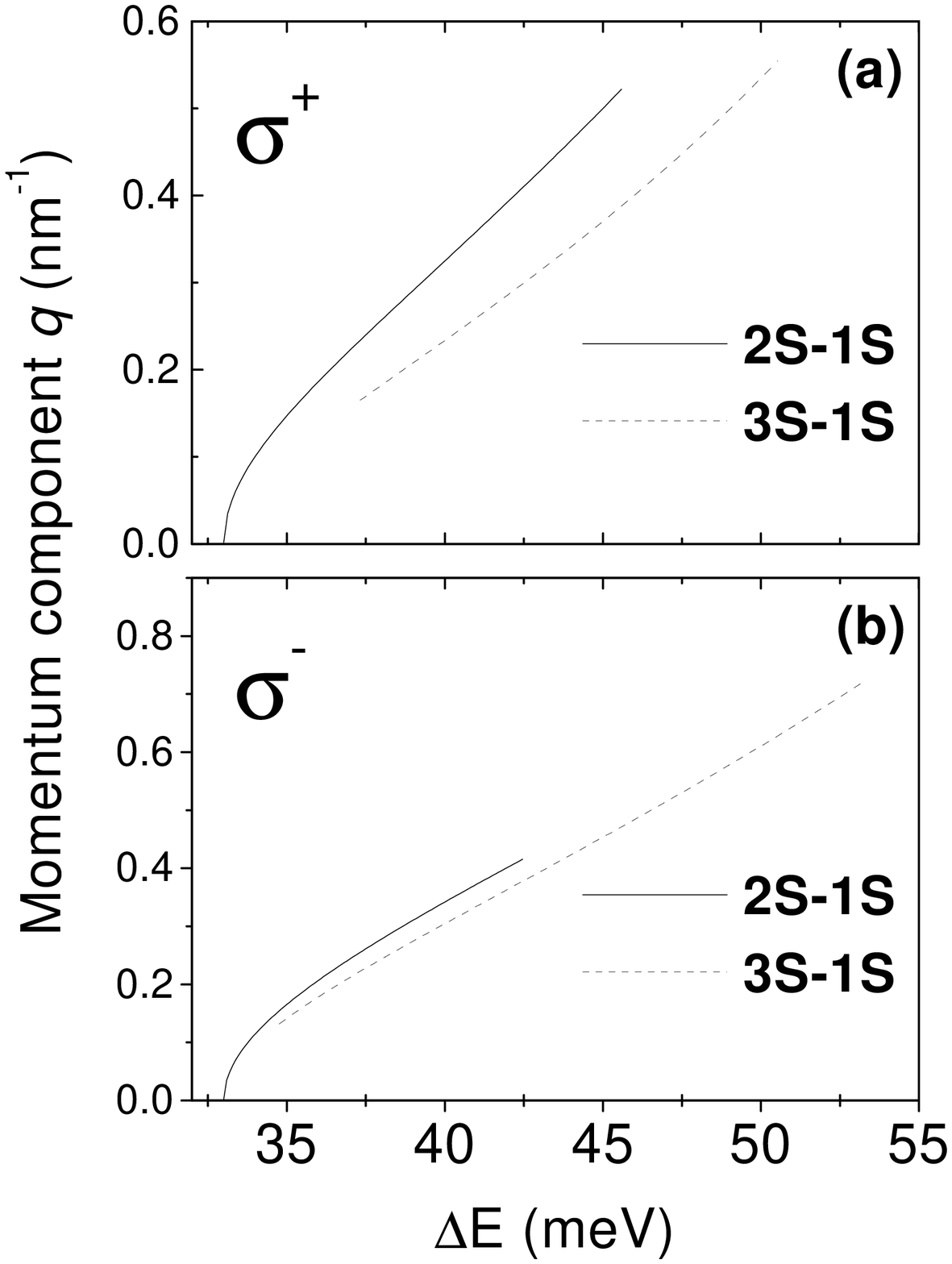}
\end{center}
\vspace{-25.mm}
 \caption{\large Values $q^{\pm}(\Delta E)$ found from the equation
$\epsilon_{\pm}(q)=\Delta E-\hbar\omega_0$ for all
transitions of $\sigma^+$ (a) and $\sigma^-$ (b) polarizations.}
\end{figure}

\begin{thebibliography}{99}
\bibitem{1sHH}
{The energy of 1sHH which we will use in the following cannot be extracted
from this set of spectra but its dependence on magnetic field was obtained
from the similar experiment previously (see Ref. [14]). In particular, for
$B=0$ this energy is $1.5805\,$meV}.
\bibitem{foot2}
{A simple estimate for the miniband width is
$\hbar^2/ML_w^2\cdot\exp{(-\alpha
L_b)}$, where the factor $\alpha$ characterizes the wave function
attenuation in the barrier which depends on the barrier height. If only
$L_w$ is changed by factor 8/3, we arrive at the indicated estimate}.
\end{thebibliography}
\end{document}